\documentclass{article}

\usepackage{arxiv}
\renewcommand{\date}[1]{}  

\usepackage[utf8]{inputenc} 
\usepackage[T1]{fontenc}    
\usepackage{hyperref}       
\usepackage{url}            
\usepackage{booktabs}       
\usepackage{amsfonts}       
\usepackage{nicefrac}       
\usepackage{microtype}      
\usepackage{lipsum}
\usepackage{graphicx}
\usepackage{amsmath}
\usepackage{amssymb}
\usepackage{cite}
\usepackage{amsmath,amssymb,amsfonts}
\usepackage{algorithmic}
\usepackage{graphicx}
\usepackage{float}
\usepackage{makecell}
\usepackage{textcomp}
\usepackage[TS1,T1]{fontenc}
\usepackage{array, booktabs}
\usepackage{mathtools} 
\usepackage{lipsum} 
\usepackage{caption}
\usepackage{comment}
\usepackage{subcaption}
\usepackage{upgreek}
\usepackage{siunitx}
\usepackage[euler]{textgreek}

\usepackage{caption}
\usepackage[super]{nth}
\usepackage[utf8]{inputenc} 

\usepackage{fancyhdr}       
\graphicspath{{media/}}     

\title{Optimisation of Magnetic Field Sensing with Optically Pumped Magnetometers for Magnetic Detection Electrical Impedance Tomography
}

\author{
  Kai Mason, Florencia Maurino-Alperovich, Kirill Aristovich and David Holder \\
  Department of Medical Physics and Biomedical Engineering \\
  University College London \\
  London, United Kingdom \\
  \texttt{kai.mason@ucl.ac.uk}
}
\begin{document}

\maketitle
\begin{abstract}
Magnetic Detection Electrical Impedance Tomography is a novel technique that could enable non-invasive imaging of fast neural activity in the brain. However, commercial magnetometers are not suited to its technical requirements. Computational modelling was used to determine the optimal number, size and orientation of magnetometers, to inform the future development of MDEIT-specific magnetometers. Images were reconstructed using three sensing axes, arrays of 16 to 160 magnetometers, and cell sizes ranging from 1 to 18 mm. Image quality was evaluated visually and with the weighted spatial variance.  Single-axis measurements normal to the surface provided the best image quality, and image quality increased with an increase in sensor number and size. This study can inform future OPM design, showing the size of the vapour cell need not be constrained to that of commercially available OPMs, and that a small array of single-axis, highly sensitive sensors is optimal for MDEIT.
\end{abstract}

\section{Introduction}

Non-invasive imaging of neural activity throughout the brain with millisecond temporal and millimetre spatial resolution remains a longstanding goal in neuroimaging. Such a breakthrough would transform both medicine and neuroscience by enabling precise monitoring of circuit activity for neurological disorders as well as resting state and evoked activity in the healthy brain. The current gold standard technique for neuroimaging is functional magnetic resonance imaging (fMRI), which typically images haemodynamic activity over seconds \cite{Logothetis2008}. Whilst fMRI has proved invaluable for applications including clinical diagnosis/monitoring and behavioural neuroscience, it does not capture the faster electrical activity of neurons, which occurs over milliseconds. Instead, this activity must be inferred from haemodynamics \cite{Logothetis2008}. 

Magnetic detection electrical impedance tomography (MDEIT) is a novel imaging technique for imaging the electrical impedance changes associated with neuronal depolarisation. It builds on the principles of electrical impedance tomography (EIT) and magnetometry, as employed in magnetoencephalography (MEG) \cite{Holder2021, Hansen2010, Mason2023a, Mason2023b}.  Recent computational modelling by Mason \textit{et al}. suggests that MDEIT could achieve a spatiotemporal resolution of approximately 1 mm and 1 ms in the human cortex. With future hardware innovations, this resolution could extend to the entire brain \cite{Mason2023b}. Achieving this capability requires optically pumped magnetometers (OPMs) for on-scalp magnetic sensing. These must have a sensitivity of $\leq10$ fTHz$^{-1/2}$ and a bandwidth of $\geq1$ kHz \cite{Mason2023a, Mason2023b}. However, OPMs meeting these specifications are not yet commercially available. Current designs focus on detecting lower-frequency ($\leq 100$ Hz) and lower-amplitude ($\sim$ 10 fT) MEG signals. This work aims to guide the future development of MDEIT-specific OPMs. It explores how the number, size, and orientation of sensors affect neural imaging performance and discusses implications for the future of the technology.

\subsection{Background}

\subsubsection{MDEIT Principle of Operation}

MDEIT is performed by injecting a constant alternating current (AC) between pairs of electrodes on the boundary of a region of interest. The resulting magnetic field is measured with magnetometers exterior to the volume \cite{Ireland2004}. Local, interior changes in the electrical impedance can be sensed as temporal changes in the amplitude of the measured AC magnetic field. By using hundreds or thousands of multiplexed electrode-magnetometer pairs, these magnetic field changes can be combined with a biophysical model to reconstruct two- or three-dimensional functional images of the underlying activity \cite{Mason2023a, Mason2023b}. 

MDEIT is an adaptation of EIT, which instead measures boundary voltages rather than exterior magnetic fields \cite{Holder2021}. A novel application of MDEIT is imaging the approximately 1 \% local impedance change associated with neuronal depolarisation in the brain in response to stimuli \cite{Mason2023b}. Previous studies have used EIT with epicortical electrodes on the rat brain to image cortical activity with a spatiotemporal resolution of 200 \textmu m and 2 ms \cite{Faulkner2018}. However, the depth resolution of fast neural EIT in the brain is limited by signal-to-noise ratio (SNR), the ill-posed nature, and rank deficiency of the EIT inverse problem. As a result, prior attempts to use scalp electrodes for fast neural EIT have been unsuccessful \cite{Holder2021, Gilad2009}.  

MDEIT offers improvements over EIT and may enable fast neural imaging with scalp electrodes. This is because (a) advancements in magnetometry could reduce noise and increase the SNR of MDEIT and (b) the MDEIT inverse problem is less rank-deficient than that of EIT, leading to higher-quality images \cite{Mason2023b}. By enabling the use of scalp electrodes instead of cortical ones, MDEIT could provide non-invasive imaging, significantly improving usability.

The AC amplitude used in fast neural MDEIT is approximately 1 mA at a frequency of $\sim 1.5$ kHz. The resulting changes in the magnetic field range from $\sim 1$ fT to $\sim 1$ pT, within a standing field of up to $\sim 10$ nT \cite{Mason2023b, Faulkner2018b}. Magnetometers for this application must have a sensitivity of $\leq 10$ fTHz$^{-1/2}$, a bandwidth of at least $\pm 500$ Hz centred at $\sim 1.5$ kHz, and a dynamic range of up to $\sim 10$ nT \cite{Mason2023b}. Currently, only superconducting quantum interference device (SQUID) magnetometers meet these specifications \cite{Neuromag2008, Rainer2019, Huber2001, Fedele2015, Storm2017, Storm2016, Koerber2023, Kado1999}. Despite their capabilities, SQUID magnetometers are expensive, require liquid helium cooling, and are fixed in place \cite{Wakai2014, Morales2017, Neuromag2008}. They also have a significant offset from the subject's head and require a magnetically shielded room \cite{Wakai2014}. These limitations present major challenges for MDEIT, particularly as movement of the current injection wires, electrodes, or the subject's head introduces artefacts that may overwhelm the impedance change signal \cite{mason2024}.  

A potential solution is a fully on-scalp system for current injection and magnetic field measurement. This would mitigate movement-related artefacts. While on-scalp current injection can be achieved using a battery-powered current source \cite{Ravagli2022}, integrating SQUID magnetometers into such a system is impractical. Optically pumped magnetometers (OPMs) could provide a feasible alternative, though they currently cannot match the performance of SQUIDs \cite{Tierney2019}.

\subsubsection{MDEIT Sensing Axes}

Fast neural MDEIT is an ill-posed and ill-conditioned problem \cite{Holder2021, Mason2023b, hansen2010discrete}. In principle, increasing the number of magnetometers should increase the image quality. However, this depends on the rank deficiency of the Jacobian (also known as the sensitivity matrix) used for image reconstruction, which measures the degree of independence between measurements \cite{Mason2023b}.

For MEG, it was recently discovered that increasing the number of OPMs above 70 provided little extra benefit for source localisation and it is expected that similar results will be obtained for MDEIT \cite{Bezsudnova2022}. Optimisation of the number of magnetometers for fast neural MDEIT may aid in the practical application of MDEIT by maximising the reconstructed image quality with limited resources. The magnetic field measured with MDEIT is a vector quantity with three orthogonal components in Cartesian space. Depending on the magnetometer; sensing of one, two or three axes may be achieved simultaneously \cite{Quspin, Kado1999, Storm2016, FieldLine, mag4health}. This begs the question of which, if any, of the axes should be prioritised if only a subset of the axes can be measured. Some magnetometers such as the QuSpin Gen-3 OPM can be operated with one, two or three-axis measurements; however, measurement with three axes comes at the cost of sensitivity. In cases such as this, it is beneficial to know the optimal mode of operation. Similarly, if only one axis may be measured then it is beneficial to know the optimal orientation of the sensor. This may also inform the design of future OPMs dedicated to MDEIT applications.

\subsubsection{OPMs}

Recent progress in OPMs has allowed for the commercial development of small, portable, room temperature OPMs with a sensitivity of $\sim 10$ ftHz$^{-1/2}$ \cite{Tierney2019, Hill2020, Borna2017, Fabricant2023, Quspin, mag4health, Morales2017}. These OPMs can be incorporated as part of an on-scalp system, cost significantly less than SQUID magnetometers and require no liquid helium, making them ideal candidates for future MDEIT systems \cite{Hill2020, Wakai2014, Tierney2019, Fabricant2023}. These advantages have led to the development of novel OPM-MEG systems which have improved sensitivity and spatial resolution compared to traditional SQUID-MEG systems \cite{Brookes2022}.

Modern commercial OPMs utilise the magnetically sensitive state of a glass cell of vapour, which is optically pumped into a quantum state using a laser. Once this is achieved, the laser light will no longer be absorbed by the vapour but will pass through to a photodiode placed on the opposite side of the cell \cite{Tierney2019}. A magnetic field in the cell induces a change in the quantum state, the laser must re-establish the quantum state via optical pumping which can be read out as a change in the voltage on the photodiode. This voltage corresponds to the magnetic field inside the cell \cite{Tierney2019}. Some OPMs may use a separate beam for probing and pumping the vapour in which the polarisation of the probe beam is altered by the magnetic field \cite{Colombo2016}, but in all cases, the magnetic field inside the cell is measured \cite{Tierney2019}. There are two common choices for the vapour inside the glass cell, alkali metals in the spin-exchange-relaxation-free (SERF) regime \cite{Alem2023, Grosz2017, Quspin, FieldLine} and metastable $^4$He \cite{mag4health, Morales2017, Labyt2019} each of which is accompanied by a different probe and pump setup.

The noise of an OPM, $N_{\text{tot}}$, can be separated into contributions from the environment (including classical electronic noise), $N_{\text{env}}$, and those intrinsic to the sensor, $N_{\text{sens}}$ \cite{Bezsudnova2022}. The relationship between these quantities is

\begin{equation}
    N_{\text{tot}} = \sqrt{N_{\text{env}}^2 + N_{\text{sens}}^2}.
\end{equation}
The environmental noise is usually attenuated using a magnetically shielded room (MSR) which can reduce this to values of $\sim 0.1$ fTHz$^{-1/2}$ \cite{Storm2017}, although this is likely higher once the current source noise of MDEIT is taken into account \cite{Mason2023b}. For MDEIT, it is feasible that this noise can be reduced such that the overwhelming majority of the noise is intrinsic to the sensor \cite{Mason2023b}. The intrinsic sensor noise can be expressed as a combination of the atomic-shot noise $N_{\text{at}}$ and the photon-shot noise $N_{\text{ph}}$ as \cite{Bezsudnova2022}

\begin{equation}
    N_{\text{sens}} = \sqrt{N_{\text{at}}^2 + N_{\text{ph}}^2}.
\end{equation}

The photon-shot noise is attributable to the quantum uncertainty of measuring light properties and the atomic-shot noise is attributable to the quantum uncertainty in the measurement of charge carriers \cite{Budker2004, Bezsudnova2022}. In an optimal configuration, the photon-shot noise does not exceed the atomic-shot noise \cite{Budker2004, Auzinsh2004} which gives a fundamental sensitivity $\delta B_{\text{opt}}$ of

\begin{equation}
\label{eq: OPM noise limit}
    \delta B_{\text{opt}} = \frac{\sqrt{2}\text{BW}}{\gamma\sqrt{nV}}
\end{equation}
where $\gamma$ is the gyromagnetic ratio of the atoms comprising the atomic vapour, $V$ is the volume of the atomic vapour cell, $n$ is the atomic density in the cell and $BW$ is the bandwidth \cite{Budker2004, Tierney2019, Allred2002, Rutkowski2014}. The bandwidth of an OPM can be expressed as 
\begin{equation}
    \text{BW} = \frac{1}{\tau},
\end{equation}
where $\tau$ is the relaxation time of the atoms in the vapour cell. This demonstrates that there is a tradeoff between bandwidth and sensitivity \cite{Tierney2019}. For modern commercial SERF OPMs with alkali vapour, these parameters may be tuned such that $\delta B_{\text{opt}} \simeq 10$ fTHz$^{-1/2}$, BW $\simeq 100$ Hz, $V \simeq 27$ mm$^3$ and $n \simeq 1.5\times10^{11}$ mm$^{-3}$ \cite{Tierney2019, Shah2013, Shah2018, Rea2022}. For commercial helium OPMs, the parameters may be tuned such that  $B_{\text{opt}} \simeq 40$ fTHz$^{-1/2}$, BW $\simeq 2000$ Hz, $V \simeq 785$ mm$^3$ and $n \simeq 1\times10^{8}$ mm$^{-3}$ \cite{Fourcault2021}. 

In practice, the performance of alkali and $^4$He magnetometers are heavily affected by laser and environmental noise; however, it is expected that these noise sources will be reduced with further technological developments and OPMs will approach their fundamental sensitivity in the future \cite{Krzyzewski2019, Fourcault2021}.

It has been determined that for both SERF and $^4$He OPMs, larger cell size allows for greater sensitivity and lower temperature of operation (for heated SERF OPM cells) \cite{Shah2007, Rutkowski2014}. For MEG, increasing the cell volume has drawbacks; the effective standoff distance between the scalp and cell is increased, the operating power is increased and the spatial resolution is decreased because of higher spatial averaging of the MEG signal \cite{Fabricant2023, Bezsudnova2022, Shah2007}. The first two of these drawbacks are certainly shared by MDEIT. However, it is not known whether the latter is equally as significant for MDEIT as MEG. For MDEIT, neither of the aforementioned commercial OPMs can achieve the necessary bandwidth and sensitivity requirements for fast neural imaging. In the case of alkali OPMs the sensitivity is comparable to that of commercial SQUID magnetometer systems but the bandwidth is insufficient. For $^4$He OPMs, the bandwidth is suitable but the noise is four times larger than that of alkali OPMs \cite{mag4health}. In either case, it is imperative that the effect of vapour cell size on the SNR magnitude and reconstructed image quality is known since this may inform future OPM design, potentially allowing for more sensitive OPMs and/or OPMs with larger bandwidths. A recent study by Bezsudnova \textit{et al}. \cite{Bezsudnova2022} found that cell vapour sizes significantly larger than those currently used could be tolerated for MEG whilst maintaining spatial resolution.

\subsection{Purpose}

This was a modelling study in which we set out to optimise the sensing axis, number and volume of optically pumped magnetometers for use with MDEIT, and to inform the future design of OPMs for MDEIT. The purpose of this work was to answer the following questions

\begin{enumerate}
    \item What is the optimal sensing axis arrangement for OPMs with MDEIT?

    \item What is the optimal number of OPMs for use in an on-scalp MDEIT system?

    \item What is the optimal cell volume for OPMs with MDEIT?

    \item What are the implications for future OPM design?
\end{enumerate}

\subsection{Experimental Design}

\subsubsection{Model}

An accurate FEM of the human head, comprising seven tissue types, was used to minimise the errors associated with model simplification. This was because the effect of the skull's electrical impedance was critical in assessing and comparing MDEIT with EIT. The perturbations considered were chosen to accurately represent the impedance change associated with fast neural activity in the human brain. The magnitude of the local impedance change was 1 \% of the grey and white matter tissue \cite{Gilad2007b, Gilad2009}, and the volume was 3.86 cm$^{3}$ which approximately corresponds to the volume of neurons activated by a visual or sensory-evoked response \cite{Pastor2003, Gilad2009}. The same volume perturbation was considered at four depths in areas broadly corresponding to the cortex, cingulate gyrus, thalamus and pons \cite{Nowinski2011}. A local impedance change of 1 \% was chosen because this was the best estimate of the change based on current evidence  \cite{Liston2012, Faulkner2018b}. 

\subsubsection{Injection Protocol}

The injection protocol which maximised the current density in the brain (grey and white matter) was chosen since, in practice, the location of the activity to be imaged will not be known, meaning a perturbation-specific injection protocol is not likely possible \cite{Faulkner2017}.  

\subsubsection{Image Reconstruction Mesh}

To avoid an `inverse crime' and improve computational implementation time, the tetrahedral mesh used for the forward problem was down-sampled to a hexahedral mesh for image reconstruction \cite{Lionheart2004}. This is common practice in the field of EIT and increases the robustness of the inverse modelling study \cite{Aristovich2016, Holder2021, Jehl2016}. 

\subsubsection{Number of Magnetometers}

Arrays of magnetometers between 16 and 160 magnetometers were considered in this work for two reasons. (1) Because this was deemed to be a realistic range that would be possible in practice. (2) 160 magnetometers was the maximum number of magnetometers for which Jacobian calculation and image reconstruction were practically viable with the available computational resources.

\subsubsection{Noise}

For optimisation of the magnetometer sensing axes and the number of magnetometers, the noise was chosen to represent realistic measurement noise, currently achievable with commercial magnetometers (such as the QuSpin OPM \cite{Quspin}) and an open-source EIT system (the ScouseTom \cite{Avery2017}). This was done so that the results from this study would be directly applicable to experimental studies, maximising the chance of successful imaging \textit{in vivo}.

For optimisation of the OPM sensing cell, the noise was chosen to comprise the OPM noise (atomic shot noise), environmental noise and current source noise. This was chosen to represent realistic measurement noise for each magnetometer size. The OPM noise was equated to the atomic shot noise because this is the dominant source of noise for optimally functioning OPMs \cite{Budker2004, Auzinsh2004}. 
\section{Methods}

\subsection{Computational Model}

\begin{figure}[ht!]
    \centering
    \includegraphics[width = 0.8\textwidth]{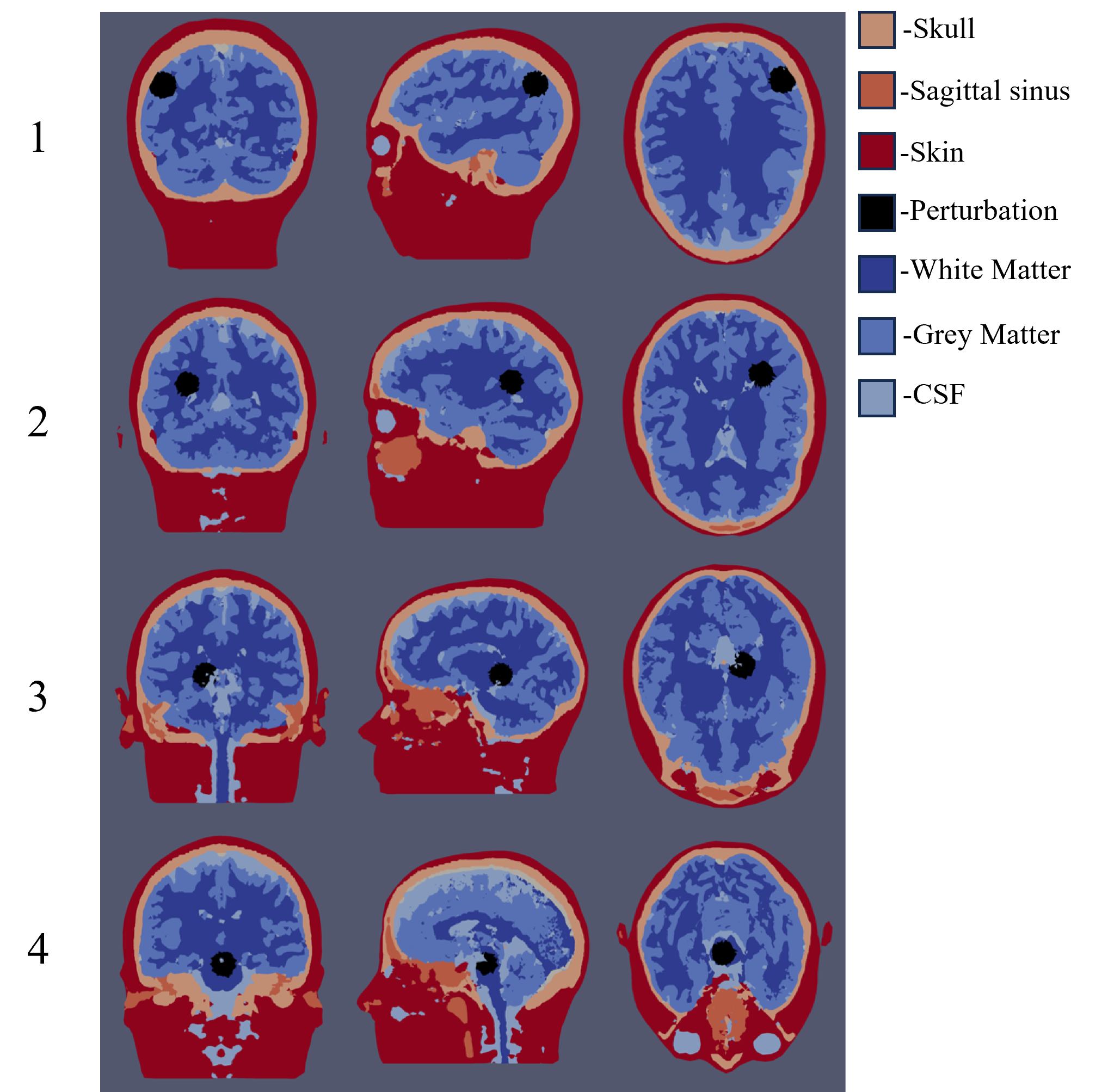}
    \caption{\textbf{A graphic representation of the FEM used in this work showing the four perturbations.} Each slice was taken through the centre of mass of the perturbation in each case. Reprinted from \cite{Mason2023b}.}
    \label{fig: FEM graphic}
\end{figure}

The computational model, perturbations and noise were almost identical to that used by Mason \textit{et al.} in \cite{Mason2023b} (Figure \ref{fig: FEM graphic}). An anatomically realistic finite element model (FEM) of a human head, comprising seven tissue types and an array of OPMs inside a cubic region of air was used for the calculation of the Jacobian and for solving the forward problem which was done using COMSOL Multiphysics \cite{comsol}. The FEM comprised 4.5 M tetrahedral elements. 32 electrodes of 5 mm radius were positioned on the scalp of the FEM in the EEG 10-20 positions \cite{Jasper1958}. The injection protocol was chosen such that the current density was maximised in the brain (grey matter and white matter) and the injection amplitude was 1 mA \cite{Faulkner2017}. 

Four approximately spherical perturbations (identical to those in \cite{Mason2023b}) of 3.86 cm$^3$ volume and 1\% local increase in conductivity were considered at depths of 7.40 mm, 32.8 mm, 58.2 mm and 83.6 mm from the inside of the skull to the centre of mass of the perturbation. The perturbations only included elements which corresponded to grey or white matter and were labelled `Perturbation 1' to `Perturbation 4' from most superficial to deepest (Figure \ref{fig: FEM graphic} and Table \ref{tab: head conductivities}). 

\begin{table}[htb]
\centering
    \begin{tabular}{c c}
    \hline
        \textbf{Tissue}  & \textbf{Conductivity (Sm$^{-1}$)}  \\
        \hline 
        White matter & 0.150 \\
        Grey matter & 0.300 \\
        Cerebrospinal fluid (CSF) & 1.79 \\
        Sagittal sinus & 0.700 \\
        Skull & 0.0180 \\
        Air & 0.0001 \\
        Scalp & 0.440 \\
        Perturbation in grey matter & 0.303 \\ 
        Perturbation in white matter & 0.1515 \\
        \hline
    \end{tabular}
    \vspace{2mm}
    \caption[Head FEM Regions of Conductivity]{\textbf{The different regions defined by the FEM and their respective conductivity} \cite{Horesh2006}. Reprinted from \cite{Mason2023b}.}
    \label{tab: head conductivities}
\end{table}

\subsection{Optimisation of Sensing Axes}
\label{sec: optimisation of sensing axes}

For optimisation of the sensing axis and number of sensing axes, 64 magnetometers were modelled as zero-dimensional points in space at a distance of 6.5 mm from the scalp. The positions of the magnetometers were generated using a Fibonacci sphere-based algorithm \cite{Mason2023b}.

At each magnetometer position, three orthogonal components of the magnetic field were computed separately and considered independent measurements. For each magnetometer, these axes were defined and calculated as:

\begin{itemize}
    \item \textbf{Axis 1: } The axis normal to the surface of the head at the closest point to the magnetometer. This was labelled $\vec{n}$ and defined as $\vec{n} = [x_1, y_1, z_1]$ in terms of the base coordinate system $\vec{x} = [x, y, z]$. 

    \item \textbf{Axis 2: } The axis tangential to the surface of the head at the closest point to the magnetometer with zero magnitude in the $z$ direction. This was labelled $\vec{t_1}$ and calculated as $\vec{t_1} = [y_1, -x_1, 0]$ where $x_1$ and $y_1$ are components of $\vec{n}$.

    \item \textbf{Axis 3: } The axis perpendicular to $n$ and $\vec{t1}$. This was labelled $\vec{t_2}$ and calculated as $\vec{t_2} = \vec{n} \times \vec{t1}$. 
\end{itemize}

The total number of measurements was $3\times64 = 192$ per injection pair; since there were 31 injection pairs, this gave a total of $5952$ measurements for each perturbation. Forward simulations and image reconstructions were performed for each axis individually and for combinations of the axes (as if multiple axes were measured simultaneously). The combinations chosen were Axis 1 (single-axis), Axes 1 and 2 together (bi-axial measurement), and all three axes together (tri-axial measurement).

Realistic noise was added to all simulated data; the values were derived from measured noise and were identical to Noise Case 1 used by Mason \textit{et al.} \cite{Mason2023b}. The uncorrelated additive noise was 316 fT due to the sensitivity of the magnetometer (which was considered to be 10 fTHz$^{-1/2}$ \cite{Quspin}) and 6.32 fT due to environmental noise \cite{Storm2017}. The correlated multiplicative noise was 0.0558 \% of the standing magnetic field due to the current source \cite{Mason2023b}. 232 measurement averages were assumed with a measurement bandwidth of $\pm$ 500 Hz. This gave a total noise of 21.2 fT + 3.67 $\times 10^{-3}$ \% \cite{Mason2023b}.

The selection of the optimal measurement axis/number of measurement axes was based on the reconstructed image quality across all three noise cases and the time/computational resources required for computational implementation (i.e., Jacobian calculation and image reconstruction). 

\subsection{Optimisation of the Number of Magnetometers}
\label{sec: optimisation of number of magnetometers}

For optimising the number of magnetometers, ten arrays of magnetometers were considered ranging from 16 to 160 in intervals of 16. The magnetometers were modelled as zero-dimensional points 6.5 mm from the scalp of the model. The positions of the magnetometers were generated using the same method as in Section \ref{sec: optimisation of sensing axes}. The magnetic field was sensed using the optimal axis configuration, which was determined beforehand based on the results of the work outlined in Section \ref{sec: optimisation of sensing axes}. Forward simulations and image reconstruction were performed for each array of magnetometers. The same noise was used as in Section \ref{sec: optimisation of sensing axes}.

\subsection{Optimisation of OPM Vapour Cell}
\label{sec: optimisation of opm vapour cell}

For optimisation of the OPM cell size, 64 magnetometers were modelled as cubic regions of space with one face normal to the nearest point on the scalp of the head model. Five sizes of magnetometer were considered ranging linearly from 1 mm to 18 mm in side length (Table \ref{tab: opm size and noise}). The distance between the face nearest the head model and the nearest point on the scalp of the head model was kept at 6.5 mm for all simulations. The magnetometers were positioned using the same method used in Sections \ref{sec: optimisation of sensing axes} and \ref{sec: optimisation of number of magnetometers}. 

The noise for each magnetometer size was calculated using Equation \ref{eq: OPM noise limit}. The parameters were based on an example SERF OPM using $^{87}$Rb. The parameters were BW = 500 Hz, $\gamma$ = 7 HznT$^{-1}$, n = 1.5 $\times$ 10$^{11}$ mm$^{-3}$ for a cubic cell size of 27 mm$^{3}$ (3 mm side length of cubic cell) \cite{Shah2009, Tierney2019}. This gave a noise of 50.2 fTHz$^{-1/2}$ for this cell size. The noise for each cell size considered in this study was based on this value and was scaled by the cell volume accordingly using Equation \ref{eq: OPM noise limit} (Table \ref{tab: opm size and noise}). In addition to this intrinsic magnetometer noise, environmental noise of 0.2 fTHz$^{-1/2}$ and current source noise of 0.0558 \% were included, as in Noise Case 1 in Sections \ref{sec: optimisation of sensing axes} and \ref{sec: optimisation of number of magnetometers} \cite{Mason2023b}.

\begin{table}[htb]
    \centering
    \begin{tabular}{c c c c}
    \hline
    OPM Side Length (mm) & OPM Volume (mm$^3$) & OPM Sensitivity (fTHz$^{-1/2}$) & Total Noise \\
    \hline
     1.00  & 1.00 & 261  & 17.1 fT + 3.67 $\times$10$^{-3}$ \% \\
     5.20  & 141  & 22.0  & 1.46 fT + 3.67 $\times$10$^{-3}$ \% \\
     9.50  & 857  & 8.91  & 0.598 fT + 3.67 $\times$10$^{-3}$ \% \\
     13.7  & 2570 & 5.14  & 0.338 fT + 3.67 $\times$10$^{-3}$ \% \\
     18.0  & 5830 & 3.42 & 0.225 fT + 3.67 $\times$10$^{-3}$ \% \\
     \hline
    \end{tabular}
    \vspace{2mm}
    \caption{\textbf{The cell volume, sensitivity and total noise after 232 measurement averages for each OPM considered in this work.} The total noise includes the intrinsic OPM sensitivity, environmental noise of 0.2 fTHz$^{-1/2}$ and current source noise of 0.0558 \% with a bandwidth of $\pm$ 500 Hz.}
    \label{tab: opm size and noise}
\end{table}

Forward simulations and image reconstruction were performed for all magnetometer sizes. The magnetic field was averaged over the sensing volume for the forward simulations. The rank of the Jacobian was also calculated for each magnetometer size.

\subsection{Algorithms}

All forward simulations and Jacobian calculations were performed using COMSOL Multiphysics \cite{comsol} using the methods developed in \cite{Mason2023b}.

For all magnetometer arrangements, perturbations and noise cases, 100 image reconstructions were performed using \nth{0} order Tikhonov Regularisation with uncorrelated noise-based correction (U-NBC) \cite{mason2024}. Images were reconstructed on a mesh comprising $\sim$376,800 cubic elements.

\subsection{Figure of Merit}

The reconstructed images were thresholded at 50 \% of the largest increase in conductivity and assessed based on the weighted spatial variance (WSV) of the images \cite{Javaherian2013}, defined as
\begin{equation}
    WSV = \left(\sum_{i=1}^{n}w_i((x_i - \bar{x})^2 + (y_i - \bar{y})^2 +(z_i - \bar{z}^2)) \right)^{\frac{1}{2}}
    \label{eq: WSV}
\end{equation}
where $x_i$, $y_i$ and $z_i$ are the $x$, $y$ and $z$ coordinates of the $i^{\text{th}}$ element in the FEM, $\bar{x}$, $\bar{y}$ and $\bar{z}$ are the $x$, $y$ and $z$ coordinates of the true centre of mass of the perturbation and $w_i$ is a weighting defined as
\begin{equation}
    w_i = \frac{S_iI_i^2}{\sum_{i=1}^{n}S_iI_i^2}
\end{equation}
where $S_i$ is the volume of the $i^{\text{th}}$ element of the FEM and $I_i$ is the reconstructed conductivity change of the $i^{\text{th}}$ element of the FEM \cite{Javaherian2013}. The WSV of the true, target images was calculated and used as a comparison for the reconstructed images and all WSV values were scaled such that the target WSV was equal to one. WSV values closer to those of the target images were considered superior. 100 images were reconstructed for each case and the statistical significance was assessed between cases using repeated measures ANOVA and the multiple comparison test. Tukey's honestly significant difference procedure was used for $p$-value scaling. 
\section{Results}

\subsection{Sensing Axis}

\begin{figure}[ht!]
    \centering
    \includegraphics[width=\linewidth]{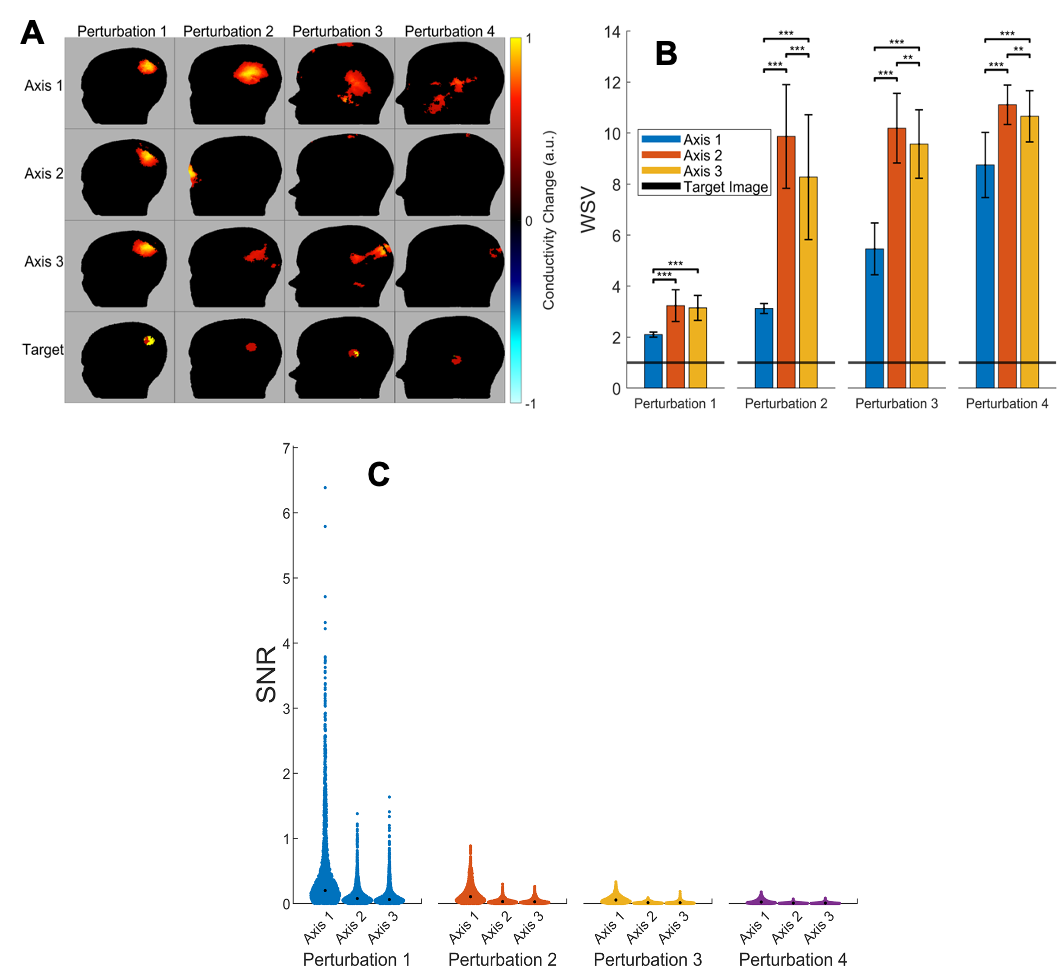}
    \caption{\textbf{The effect of the OPM sensing axis on the image quality.} (\textbf{A}) Example reconstructions for each perturbation with single-axis MDEIT. Each image is a sagittal slice of a three-dimensional reconstruction taken through the centre of mass of the perturbation. All images have been thresholded at 50 \% of the largest increase in conductivity. (\textbf{B}) The WSV (mean $\pm$ SD) for image reconstructions with single-axis MDEIT for all four perturbations. The statistical significance is shown between the measurements with each axis for each perturbation (repeated measures ANOVA and multiple comparison test, N = 100). `*' indicates that $p < 0.05$, `**' indicates that $p < 0.01$ and `***' indicates that $p < 0.001$. A WSV value closer to the black line indicates a higher-quality image reconstruction. (\textbf{C}) Swarm charts showing the distribution of the SNR of the simulated magnetic field changes due to four perturbations representing neural activity in the brain. The distribution is shown for each measurement axis of the magnetometers. Perturbation 1 is the most superficial, and Perturbation 4 is the deepest. Each point represents a separate measurement of the magnetic field change. The median SNR is shown as a black dot for each distribution. N = 1984 for each distribution. }
    \label{fig: single axis}
\end{figure}

For all perturbations, the largest maximum and median SNR was the largest with Axis 1 (Figure \ref{fig: single axis}C). On visual inspection, reconstructions with Axis 1 were of the same or superior quality to those with Axis 2 and Axis 3 with respect to the target image. The difference in quality on visual inspection between the reconstructions with each axis decreased as the noise decreased (Figure \ref{fig: single axis}A). The WSV was significantly closer to that of the target image with Axis 1 for all perturbations ($p < 0.001$, repeated measures ANOVA and multiple comparison test, N = 100).  In some cases, the image reconstruction did not correspond to the target image for reconstructions with any axis (e.g. Perturbation 4) (Figures \ref{fig: single axis}A and \ref{fig: single axis}B).

\subsection{Number of Axes}

\begin{figure}[ht!]
    \centering
    \includegraphics[width=\linewidth]{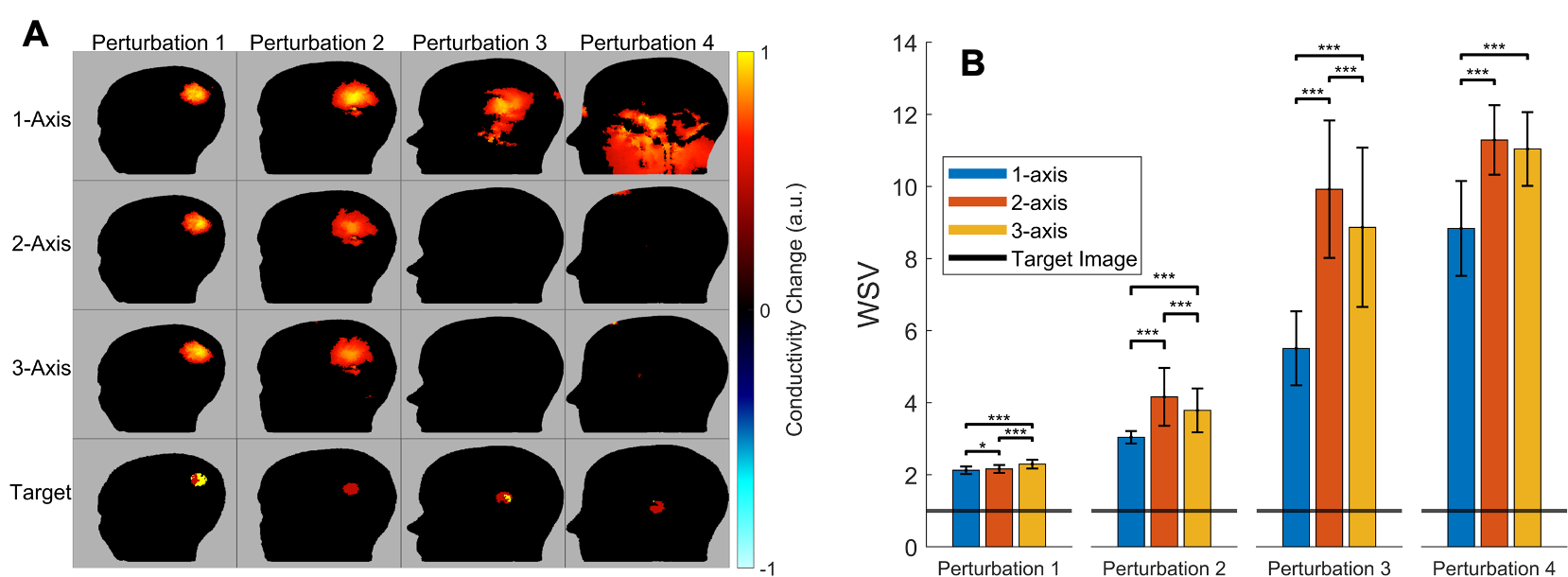}
    \caption{\textbf{The effect of the number of OPM sensing axes on the image quality.} (\textbf{A}) Example reconstructions for each perturbation with one-, two- or three-axis MDEIT. Each image is a sagittal slice of a three-dimensional reconstruction taken through the centre of mass of the perturbation. All images have been thresholded at 50 \% of the largest increase in conductivity. (\textbf{B}) The WSV (mean $\pm$ SD) for image reconstructions with 1-, 2- and 3-axis MDEIT for all four perturbations. The statistical significance is shown between the measurements with each axis for each perturbation (repeated measures ANOVA and multiple comparison test, N = 100). `*' indicates that $p < 0.05$, `**' indicates that $p < 0.01$ and `***' indicates that $p < 0.001$. A WSV value closer to the black line indicates a higher-quality image reconstruction.}
    \label{fig: multi axis}
\end{figure}

On visual inspection, 1-axis MDEIT reconstructed superior images to 2- and 3-axis MDEIT for all perturbations (Figure \ref{fig: multi axis}A). The WSV was significantly closer to that of the target image with 1-axis measurement than with 2- or 3-axis measurement for all perturbations  ($p < 0.05$, repeated measures ANOVA with multiple comparison test, N = 100) (Figure \ref{fig: multi axis}B). 

\subsection{Optimal Axis Measurement}
\label{sec: optimal axis decision}

Calculation of the Jacobian took approximately 1 day for each individual measurement axis. Reconstruction of 100 images simultaneously took approximately 5 minutes for 1-axis reconstructions, 20 minutes for 2-axis reconstructions and 60 minutes for 3-axis reconstructions. All computations used a workstation computer (CPU: AMD Ryzen Threadripper PRO 5965WX 24-Cores, RAM: 500 GB).  1-axis measurement with Axis 1 was identified as the optimal sensing axis configuration based on the computational time and image quality.

\subsection{Number of OPMs}

\begin{figure}[ht!]
    \centering
    \includegraphics[width=\linewidth]{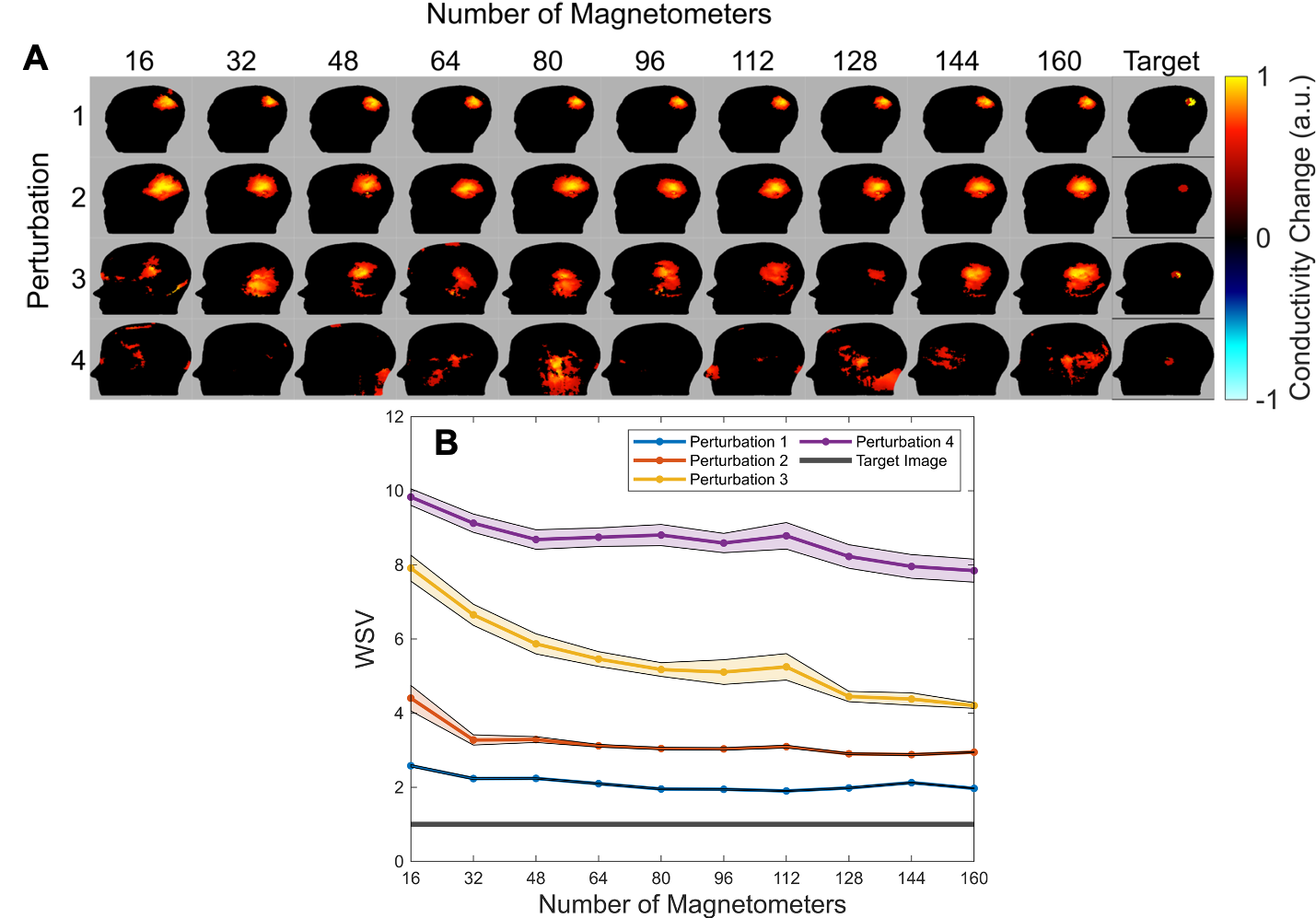}
    \caption{\textbf{The effect of the number of OPMs on the image quality.} (\textbf{A}) Example reconstructions for each perturbation with MDEIT with arrays of 16 to 160 magnetometers. Each image is a sagittal slice of a three-dimensional reconstruction taken through the centre of mass of the perturbation. All images have been thresholded at 50 \% of the largest increase in conductivity. (\textbf{B}) The WSV (mean $\pm$ 95 \% CI) for image reconstructions with MDEIT with arrays of 16 to 160 magnetometers for all four perturbations. A WSV value closer to the black line indicates a higher-quality image reconstruction.}
    \label{fig: number OPMs}
\end{figure}

A single-axis measurement with Axis 1 was used for all reconstructions as stated in Section \ref{sec: optimal axis decision}). On visual inspection, the image quality increased as the number of magnetometers increased for all noise cases (Figure \ref{fig: number OPMs}A). The WSV value approached the target WSV value as the number of magnetometers increased and the rate of change of the WSV decreased as the number of magnetometers increased (Figure \ref{fig: number OPMs}B). 

\subsection{Effect of OPM Cell Volume}
\begin{figure}[ht!]
    \centering
    \includegraphics[width=\linewidth]{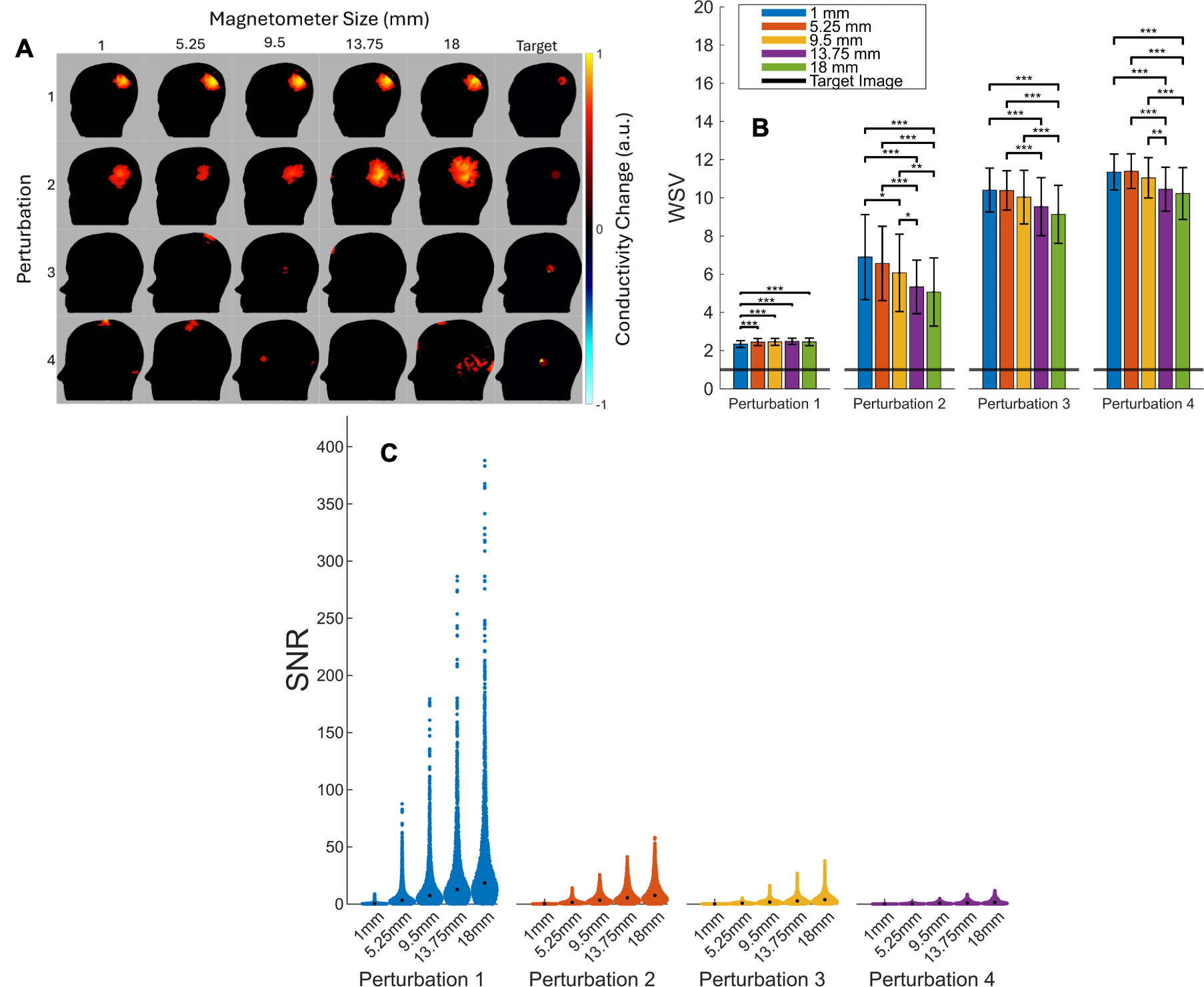}
    \caption{\textbf{The effect of the OPM sensing volume on the image quality.} (\textbf{A}) Example reconstructions for each perturbation with MDEIT with 64 magnetometers ranging from 1 to 18 mm in size. Each image is a sagittal slice of a three-dimensional reconstruction taken through the centre of mass of the perturbation. All images have been thresholded at 50 \% of the largest increase in conductivity. (\textbf{B}) The WSV (mean $\pm$ SD) for image reconstructions with single-axis MDEIT for all four perturbations. The statistical significance is shown between the measurements with each magnetometer size for each perturbation (repeated measures ANOVA and multiple comparison test, N = 100). `*' indicates that $p < 0.05$, `**' indicates that $p < 0.01$ and `***' indicates that $p < 0.001$. A WSV value closer to the black line indicates a higher-quality image reconstruction. (\textbf{C}) Swarm charts showing the distribution of the SNR of the simulated magnetic field changes due to four perturbations representing neural activity in the brain. The distribution is shown for magnetometer sizes ranging from 1 to 18 mm. Perturbation 1 is the most superficial, and Perturbation 4 is the deepest. Each point represents a separate measurement of the magnetic field change. The median SNR is shown as a black dot for each distribution. N = 1984 for each distribution.}
    \label{fig: OPM size}
\end{figure}

A single-axis measurement with Axis 1 was used for all reconstructions as stated in Section \ref{sec: optimal axis decision}. On visual inspection, the image quality increased as the magnetometer size increased for perturbations 1 and 2. Perturbations 3 and 4 failed to reconstruct regardless of magnetometer size (Figure \ref{fig: OPM size}A). The WSV value approached the target WSV value as magnetometer size increased, with the rate of improvement being most visible for perturbation 2 (Figure \ref{fig: OPM size}B). The SNR increased with increasing magnetometer size (Figure \ref{fig: OPM size}C). The rank of the Jacobian for each OPM cell volume remained constant as a function of the OPM cell size.

\section{Discussion}

\subsection{Summary of Results}

Forward and inverse MDEIT simulations were performed to optimise the sensing arrangement for fast neural MDEIT with OPMs. It was found that the measurement of the magnetic field normal to the surface of the head provided the largest SNR and best image quality; the addition of more sensing axes provided little increase in image quality. Similarly, increasing the number of magnetometers provided diminishing returns in image quality and 64 was confirmed to be a suitable number of magnetometers. Increasing the volume of the OPM vapour cell increased the SNR and reconstructed image quality. 

\subsection{Study Limitations}

This study only considered one injection protocol, which was optimised to maximise the current density inside the brain. For such a protocol, the injection pairs of electrodes were often on approximately opposite sides of the geometry. This produced approximately similar current distributions for each injection pair, which may be characteristic of the optimal axis measurement. This may limit the applicability of these results, if a different kind of injection protocol is to be used then these results should be verified for such a protocol. 

\subsection{What is the optimal sensing axis arrangement for OPMs with MDEIT?}

Analysis of the SNR and image quality showed that measurement with Axis 1 (normal to the scalp) was superior to Axes 2 and 3 (tangential to the scalp). A single-axis measurement with Axis 1 was the optimal configuration and is recommended for use in practice. The increase in image quality with Axis 1 was attributable to the larger SNR, the addition of measurements from Axes 2 and 3 added noisy measurements, which reduced the image quality. If the noise were much lower, the addition of measurements from these axes may not reduce the image quality. 

\subsection{What is the optimal number of OPMs for use in an on-scalp MDEIT system?}

Observation of the reconstructed images and the WSV showed diminishing returns in image quality as the number of magnetometers increased. For Perturbations 1 and 2, increasing the number of magnetometers above 32 provided almost no benefit to image quality; however, the image quality continued to increase up to 160 magnetometers for the deeper perturbations. 

The Jacobian calculation method used for these studies meant that increasing the number of magnetometers corresponded to an approximately linear increase in the computational time required to calculate the Jacobian \cite{Mason2023b}. In practice, there was an increase in Jacobian calculation time from $\sim$ 1 day to $\sim$ 1 week between 16 magnetometers and 160 magnetometers. Taking this into account, as well as the increased cost of using more magnetometers and potential technical issues which may be encountered in practice, an array of between 48 and 96 magnetometers is practically suitable in most cases. Increasing the number above this will provide a relatively small increase in image quality. However, using more magnetometers is always technically beneficial, and as many as are available should be used, provided that the necessary computational resources are available.

\subsection{What is the optimal cell volume for OPMs with MDEIT?}
The reconstructed images showed little visible improvement in image quality with varying OPM cell volume, particularly with perturbations 3 and 4. The WSV, however, showed a clear increase in image quality with larger cell volumes, with this effect being more pronounced for the deeper perturbations. While there could be concerns that a larger cell volume would lead to spatial averaging, and therefore reduced spatial information, the rank of the Jacobian remained the same for all cell volumes, meaning no extra unique information is expected when using a smaller OPM. 

Taking these factors into consideration, increasing the cell volume of the OPM for a given number of sensors is expected to improve the image quality.

\subsection{What are the implications for future OPM design?}

These results suggest that the primary focus for future OPM development should be to measure with one axis with as high sensitivity as possible.  If sensing with multiple axes can be achieved with the same sensitivity as a single-axis measurement, this will provide a benefit. However, this is not as important as reducing the noise for one axis. 

Increasing the number of magnetometers above approximately 128 provided little benefit to the image quality in all cases and increasing the number above 32 provided little benefit for imaging of cortical activity. This should be factored into the design of OPM arrays for MDEIT. Since cortical imaging is the most immediate objective, costs can be saved by using smaller arrays of more sensitive magnetometers. 

These results also demonstrated that, for the same number of total measurements, increasing the number of magnetometers is better than increasing the number of sensing axes. For example, the image reconstructions with 128 magnetometers were consistently of superior quality to those with two-axis measurements with 64 magnetometers. Furthermore, increasing the number of magnetometers produced higher-quality images than increasing the cell volume did. This indicates that future OPM design should focus on developing a larger number of cheaper single-axis OPMs as opposed to fewer, larger, more expensive multi-axis OPMs.

The most significant finding of this work is that the image quality increased with the size of the OPM vapour cell. This demonstrates that future OPMs for MDEIT need not be constrained to the size of current commercial OPMs optimised for MEG \cite{Quspin, mag4health}. This allows for potential increases in bandwidth and/or sensitivity which may enable fast neural MDEIT with fully on-scalp electronics, a magnetic sensitivity comparable to that achieved by SQUIDs and a bandwidth sufficient for millisecond-resolution imaging of neural activity. Future work should focus on the design and fabrication of such OPMs, particularly those dedicated to fast neural MDEIT.

\subsection{Author Contributions}
\textbf{K.M.} Conceptualisation, methodology, software, analysis, investigation, writing, visualisation. \textbf{F.M.A} Software, analysis, investigation, writing, visualisation. \textbf{K.A.} Supervision.  \textbf{D.H.} Supervision. 

\subsection{Data Availability}
The data and code presented here are available upon request to the corresponding author.

\subsection{Acknowledgements}
This work was supported by the EPSRC DTP Research Studentship [EP/N509577/1], [EP/T517793/1], [EP/X018415/1] and EPSRC UCL i4health CDT [EP/S021930/1]. 

\subsection{Conflict of Interests}
\textbf{D.H.} is the director of Cyqiq Ltd which has an interest in commercialising fast neural EIT. However, it has no current activity in MDEIT.

\clearpage

\bibliography{bibliography}
\bibliographystyle{IEEEtran}

\end{document}